%% file: cosmography_revision3_190520.tex
\documentclass[12pt]{article}

\usepackage{scicite,float}

\usepackage{times,graphicx,amsmath,tikz}
\usepackage[caption=false]{subfig}

\usepackage{multirow, makecell, booktabs}
\usepackage[capitalise,nameinlink,noabbrev]{cleveref}
\Crefformat{figure}{#2Fig.~#1#3}
\Crefmultiformat{figure}{Figs.~#2#1#3}{ and~#2#1#3}{, #2#1#3}{ and~#2#1#3}

\newcommand{\head}[1]{\textnormal{\textbf{#1}}}
\newenvironment{sciabstract}{%
\begin{quote} \bf}
{\end{quote}}

\newcounter{lastnote}

\title{A measurement of the Hubble constant from angular diameter distances to two gravitational lenses}

\author
{Inh Jee,$^{1\ast}$ Sherry H. Suyu,$^{1,2,3\ast}$ Eiichiro Komatsu,$^{1,4}$ \\Christopher D. Fassnacht,$^{5}$  Stefan Hilbert,$^{6,7}$ L\'{e}on V. E. Koopmans$^{8}$\\
\\
\normalsize{$^{1}$Max-Planck-Institut f\"ur Astrophysik}\\
\normalsize{
85741 Garching, Germany}\\
\normalsize{$^{2}$Institute of Astronomy and Astrophysics, Academia Sinica}\\
\normalsize{11F of Astronomy-Mathematics Building, No.1, Section 4, Roosevelt Road, 
Taipei 10617, Taiwan}\\
\normalsize{$^{3}$Physik-Department, Technische Universit\"at M\"unchen, 
85748 Garching, Germany}\\
\normalsize{$^{4}$Kavli Institute for the Physics and Mathematics of the Universe}\\
\normalsize{Todai Institutes for Advanced Study, the University of Tokyo, Kashiwa 277-8583, Japan}\\
\normalsize{
(Kavli IPMU, World Premier International Research Center Initiative (WPI)) 
}\\
\normalsize{$^{5}$Physics Department, University of California, Davis}\\
\normalsize{
Davis, CA 95616, USA}\\
\normalsize{$^{6}$Exzellenzcluster Universe}\\
\normalsize{
85748 Garching, Germany}\\
\normalsize{$^{7}$Ludwig-Maximilians-Universit\"at, Universit\"ats-Sternwarte}\\
\normalsize{
81679 M\"unchen, Germany}\\
\normalsize{$^{8}$Kapteyn Astronomical Institute, University of Groningen}\\
\normalsize{
9700 AV Groningen, The Netherlands}\\
\normalsize{$^\ast$To whom correspondence should be addressed;} \\
\normalsize{E-mail: jee1213@mpa-garching.mpg.de, suyu@mpa-garching.mpg.de.}\\
}

\date{}

\begin{document} 


\baselineskip24pt


\maketitle


\begin{sciabstract}
The local expansion rate of the Universe is parametrized by the Hubble constant, $H_0$, the ratio between recession velocity and distance. Different techniques lead to inconsistent estimates of $H_0$. Observations of Type Ia supernovae (SNe) can be used to measure $H_0$, but this requires an external calibrator to convert relative distances to absolute ones.
 We use the angular diameter distance to strong gravitational lenses as 
 a suitable calibrator, which is only weakly sensitive to cosmological assumptions.
 We determine the angular diameter distances to two gravitational
 lenses, $810^{+160}_{-130}$ and $1230^{+180}_{-150}$~Mpc, at redshifts of $z=0.295$
 and $0.6304$. Using these absolute distances to calibrate 740 previously-measured relative distances to SNe, we measure the Hubble constant to be
 $H_0=82.4^{+8.4}_{-8.3} ~{\rm km\,s^{-1}\,Mpc^{-1}}$.
\end{sciabstract}


Measurements of extragalactic distances have revealed that the Universe is
expanding \cite{hubble:1929}, and the expansion is accelerating
\cite{perlmutter/etal:1998, riess/etal:1998}. The distance-redshift relation is normalized using
the Hubble constant $H_0$. The value of $H_0$ has been inferred directly
from the distance-redshift relation in the local Universe, e.g.,
\cite{riess/etal:2016, freedman/etal:2012}, and indirectly from the
cosmic microwave background (CMB) by assuming a
cosmological model \cite{ade/etal:2015, Planck:2018i}. Some researchers claim that these two determinations do not agree, differing by a formal statistical significance of more than 3$\sigma$ \cite{bernal/etal:2016,freedman/etal:2017}. Various interpretations for this discrepancy have been suggested: e.g. a modification in early Universe physics  \cite{bernal/etal:2016, lemos/etal:2018}. Other researchers claim that the tension is not statistically significant: e.g.  that the tension is only at the 2.5$\sigma$ level \cite{abbott/DES:2017} or less  \cite{Zhang++2017}. $H_0$ determinations using galaxy clusters and ages of old galaxies at intermediate redshift (e.g. \cite{daSilva:2018}) are in agreement with the value from the CMB. If confirmed by further measurements, preferably using independent
methods, this discrepancy would call for a revision of the standard
model of cosmology, $\Lambda$ Cold Dark Matter ($\Lambda$CDM).

There are multiple ways to measure distances in the Universe based on our knowledge of an object whose distance is measured. One of them is the ``luminosity distance'' $D_\mathrm{L}$, which is defined based on the relationship between the measured flux $F$ and the known luminosity $L$ of an object; $D_\mathrm{L} = \sqrt{L/(4\pi F)}$. Another way to obtain distance is via the ``angular diameter distance'' $D$, where the measured angular size $\theta$ of an object is related to the known physical size of the object $r$ as $D = r/\theta$. Luminosity distances to type Ia supernovae (SNe) can be used to
determine $H_0$; however, they provide only relative distances
because of uncertainty in their absolute brightness. SNe measurements of $H_0$ must adopt an
external calibrator of the absolute distance scale, which we refer to as an anchor, to
fix the overall normalization of the distances to SNe. Existing local
distance measurements use Cepheid variable stars, parallaxes, and / or masers as anchors
\cite{riess/etal:2016}, thereby constructing a distance ladder.

We apply an independent method \cite{jee/etal:2015a} to measure angular diameter distances to 
strong gravitational lenses. We apply it to two examples at
$z=0.295$ and $0.6304$ with time-varying source brightness, where $z$ is the redshift. 
Our goal is to use the two absolute distances to anchor the relative distances of
SNe, constraining $H_0$. If we can determine
the absolute distances to $z=0.295$ and $0.6304$, we can use them to calibrate SNe data over a wider redshift range $0 <z < 1.4$. From this, we aim to determine the expansion rate at $z=0$, i.e., $H_0$. This is an inverse distance ladder method
\cite{aubourg/etal:2015,cuesta/etal:2014}.

\begin{figure}
\begin{minipage}{0.5\textwidth}
\begin{center}
\includegraphics[width=7cm]{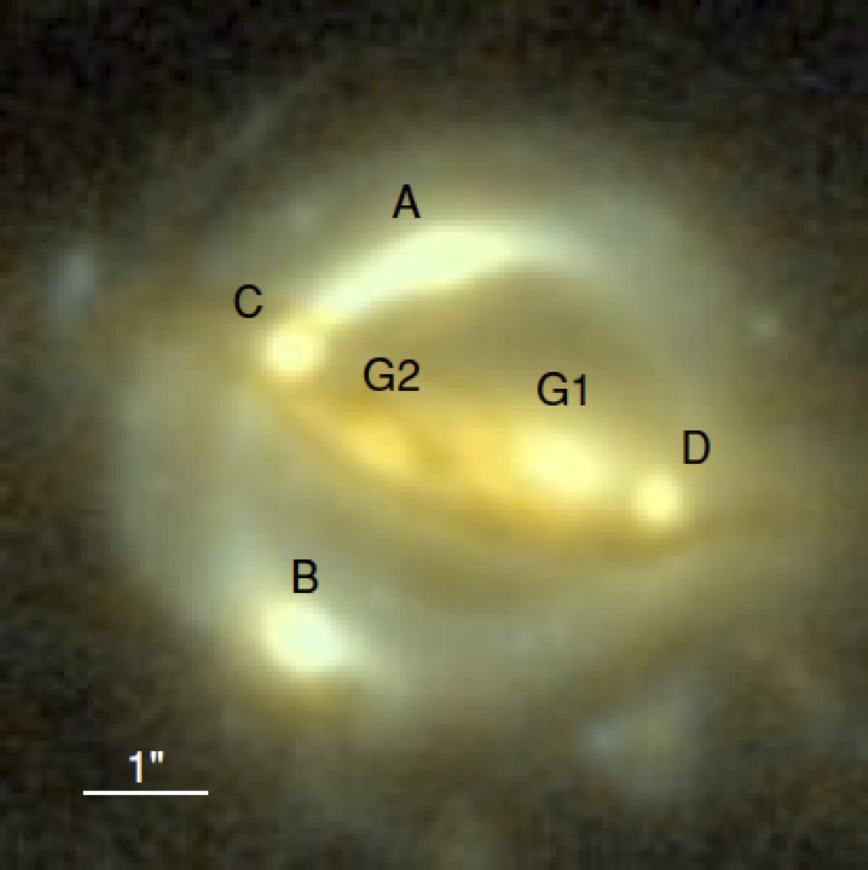}\\(A)
\end{center}
\end{minipage}
\begin{minipage}{0.5\textwidth}
\begin{center}
\includegraphics[width=7cm]{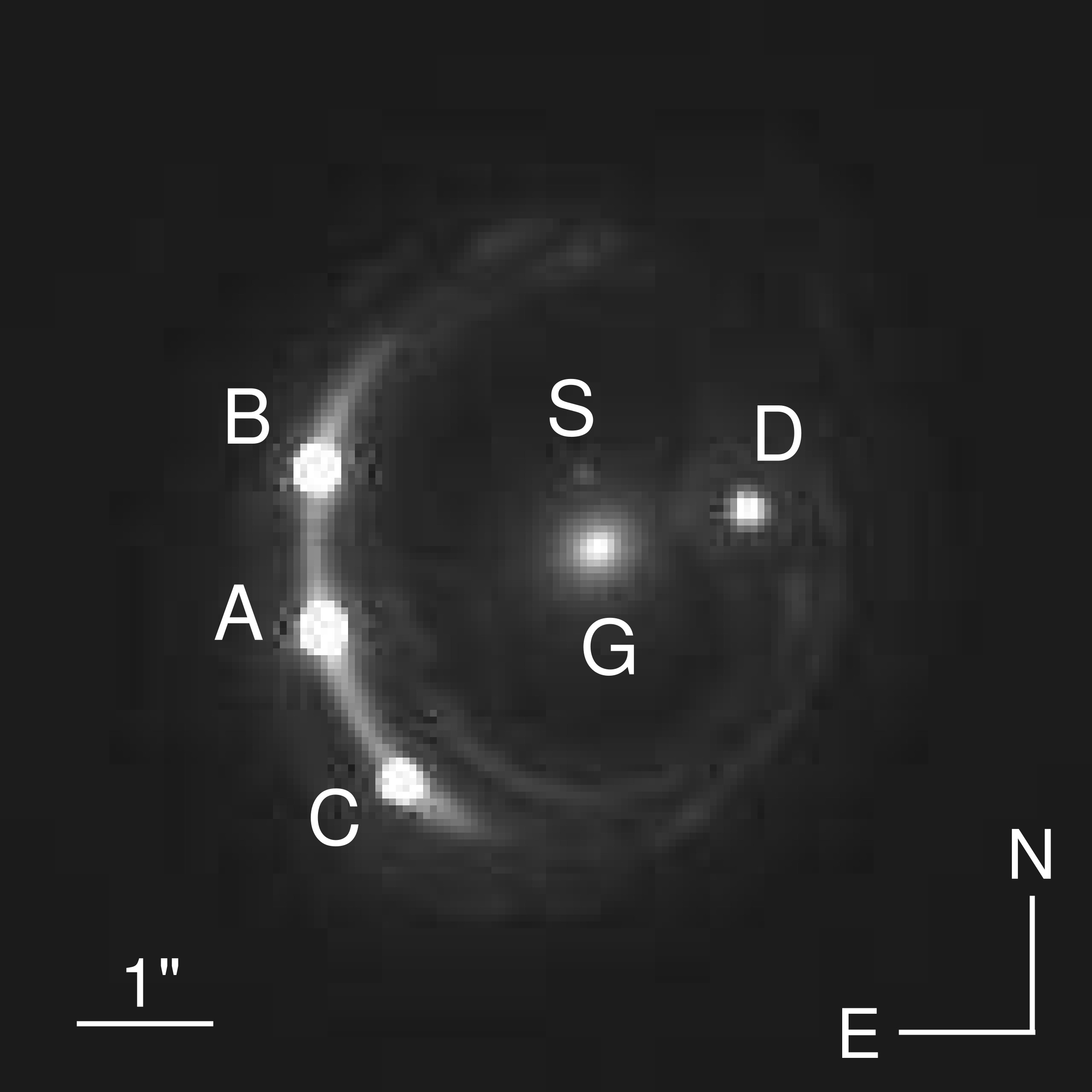}\\(B)
\end{center}
\end{minipage}
\begin{minipage}{1.0\textwidth}
\begin{center}
\includegraphics[width=16cm]{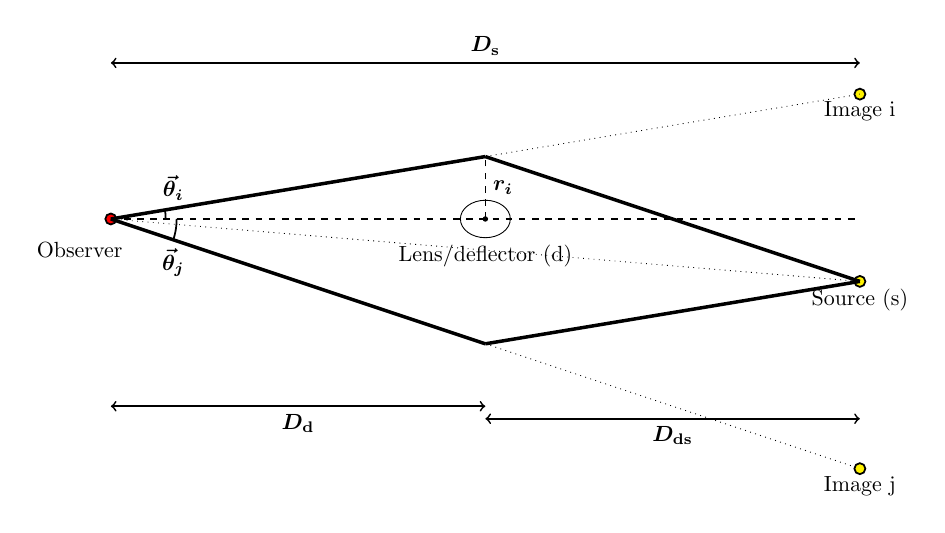}\\(C)
\end{center}
\end{minipage}
\caption{\textbf{Images of B1608+656, RXJ1131$-$1231 and the lensing configuration.} Hubble Space Telescope (HST) Advanced Camera for Surveys (ACS) F814W and F606W color composite image of lens B1608+656 (A) and ACS F814W image of RXJ1131$-$1231 (B) \cite{suyu/etal:2009,koopmans/etal:2003,sluse/etal:2003}. Labels A-D are the quasar images, G (1-2) are lens galaxies, and S is a satellite galaxy. (C) A schematic diagram of light paths from the source to the observer, forming multiple images. Lensing observables and distances are labelled, where subscripts $i$ and $j$ are the image indices. Panel (A) is reproduced from figure 1 of \cite{suyu/etal:2010} with permission.}
\label{fig:lenses}
\end{figure}

Gravitational lensing occurs when photons emitted from a background source are deflected by the
gravitational potential of a foreground massive object, such as a galaxy. An observer sees photons arriving from different directions at different times in the case of strong lensing, and these form multiple images on the sky. We show images of the two lensing systems, CLASS\,B1608+656 \cite{suyu/etal:2009,koopmans/etal:2003,fassnacht/etal:2002} (hereafter B1608+656)
and RXS\,J1131$-$1231
\cite{suyu/etal:2012,tewes/etal:2013,sluse/etal:2003} (hereafter RXJ1131$-$1231) in \cref{fig:lenses} A-B, and schematics of the system configurations in \cref{fig:lenses} C. The foreground galaxy that dominates the deflection is defined as the main lens, while the
deflections caused by any other structure along the line of sight external to
the lens is parameterized by the external convergence,
$\kappa_{\rm ext}$.

When the source brightness is variable, the arrival time difference between photons from
different images (i.e. the time delay) can be measured. Physically, two effects contribute to the difference in photon arrival time: the projected gravitational potential of the enclosed mass, and the difference in geometric path lengths between images, which are summarized as the Fermat potential $\phi$. The
time delay, $\Delta t$, between two images is given by
$c\Delta t=D_{\Delta t}\Delta\phi$ \cite{refsdal:1964}, where $D_{\Delta t}$ is
the time-delay distance $D_{\Delta t} =
(1+z_\mathrm{d})D_\mathrm{d} D_\mathrm{s}/{D_\mathrm{ds}}$, $c$ is the speed of light, $\Delta \phi$ is Fermat potential difference between the two images,
$z_\mathrm{d}$ is the lens / deflector redshift, $D$ is the angular diameter
distance, and subscripts `d' and `s' stand for the deflector and the
source, respectively (thus $D_\mathrm{ds}$ is the angular diameter distance between the deflector and the source). The time-delay distance thus relates $\phi$ to $\Delta t$. 
The external convergence, $\kappa_\mathrm{ext}$, modifies the
relationship between the true $D_{\Delta t}$ and the observed $\Delta t$
as $c\Delta t = (1-\kappa_{\rm ext})D_{\Delta t}\Delta\phi$, where $\Delta \phi$ 
is the Fermat potential difference based on a model that does not account for the external convergence.  
Therefore, the true $D_{\Delta t}$ will be scaled by $1/(1-\kappa_\mathrm{ext})$ 
for an observed (fixed) $\Delta t$. 
Several measurements of $H_0$ have been reported using $D_{\Delta t}$ alone, which scales inversely to $H_0$ and weakly depends on other cosmological parameters \cite{suyu/etal:2012, bonvin/etal:2016}.
The latest
determination yields $H_0=71.9^{+2.4}_{-3.0}~{\rm km\,s^{-1}\,Mpc^{-1}}$, which
agrees with $H_0=73.48\pm 1.66~{\rm
km\,s^{-1}\,Mpc^{-1}}$ from the local distance ladder method \cite{riess/etal:2018},
but is higher than the CMB result assuming a flat $\Lambda$CDM model,
$H_0=67.4\pm 0.5~{\rm km\,s^{-1}\,Mpc^{-1}}$ 
\cite{planck:2018vi}. All uncertainties are the 68\% confidence interval.

It is possible to measure the angular diameter distance to the deflector, $D_\mathrm{d}$, directly using a simple spherical lens model which relates the radial mass density profile $\rho(r)$ to a radius-independent velocity dispersion $\sigma^2$ following $\rho(r)=\sigma^2/(2\pi Gr^2)$  \cite{paraficz/hjorth:2009}, where $G$ is the gravitational constant. The time-delay difference between two images in this model is
given by $\Delta t=D_{\Delta t}(\theta_1^2-\theta_2^2)/(2c)$ where
$\theta_1$ and $\theta_2$ are angular positions of the two images (as illustrated in \cref{fig:lenses} C). The image
positions are related to the velocity dispersion as
$\sigma^2=[(\theta_1+\theta_2)c^2/8\pi]D_\mathrm{s}/D_\mathrm{ds}$. Combining
the two, we obtain $D_\mathrm{d}=c^3\Delta
t/[4\pi\sigma^2(1+z_\mathrm{d})\Delta\theta]$ with $\Delta\theta\equiv
\theta_1-\theta_2$ \cite{paraficz/hjorth:2009}.  This expression allows us to determine $D_\mathrm{d}$ from measurements of $\Delta t$, $\sigma$ and $\Delta \theta$. Similar, but more complex, relations hold for more generic lenses with different density profile and velocity structure 
\cite{jee/etal:2015a}.

The scaling of $D_\mathrm{d}$ with $\Delta t$, $\sigma^2$ and $\Delta\theta$ can be demonstrated by a qualitative argument. The time delay constrains the projected gravitational potential of the
lens within some characteristic size of the system $r$ (e.g. the effective radius of the lens galaxy, where half of the total light emitted from the galaxy is contained) and mass $M$, $\Delta t\sim GM \ln (r)$, while
the velocity dispersion of stars in the lens galaxy, $\sigma^2$, constrains the
gravitational potential of the lens, $\sigma^2\sim GM/r$. By combining
the two, $r$ is constrained, and by
comparing $r$ to the angular separation of
lensed images $\Delta\theta$, the lens effectively becomes a ruler, allowing the angular
diameter distance to the lens $D_\mathrm{d} =r/\Delta\theta\sim \Delta t/(\sigma^2\Delta\theta)$ to be
obtained. The physical interpretation of $r$ depends on the lens mass distribution. We adopt the modelling of the lens mass distribution and source light for both of these lensing systems \cite{suyu/etal:2009, suyu/etal:2012, suyu/etal:2010}; this allows us to use $\Delta t$ and $\sigma$ from observations but model the full surface brightness distribution of the lensed source (instead of $\Delta \theta$) in determining $D_\mathrm{d}$. The inference of $H_0$ from $D_{\rm d}$ is independent of $\kappa_{\rm ext}$, in contrast to the $H_0$ inference from $D_{\Delta t}$ that is scaled by $1/(1-\kappa_{\rm ext})$ \cite{jee/etal:2015a, suyu/etal:2013}.

As $D_\mathrm{d}\sim \Delta t/(\sigma^2\Delta \theta)$, the uncertainty on
$D_\mathrm{d}$ is determined by those on $\Delta t$, $\sigma ^2$ and $\Delta \theta$; the uncertainty in $\sigma^2$
dominates \cite{jee/etal:2015a}. $GM/r$ is determined by the
radial component of the stellar velocity dispersion, which is not
observable directly. We must assume a 3-dimensional structure for
the velocity dispersion to relate the observable line-of-sight
$\sigma^2$ to just the radial component. This velocity anisotropy
is the dominant source of uncertainty in this method \cite{jee/etal:2015a}.


Published observations of the lens galaxies provide the velocity
dispersion averaged over an aperture of a fixed physical size, which we refer to as the kinematics data. The
velocity dispersion of RXJ1131$-$1231 is estimated via spectroscopy with
a rectangular aperture of area $0.81''\times 0.7''$, where the center
of the aperture is placed at the center of the lens galaxy. The
effective radius of the lens galaxy is $R_\mathrm{eff} =
1.85''$; thus, the half width of the aperture is $\sim$20\% of the effective radius. For B1608+656, $R_{\rm eff} = 0.58''$, while $R_{\rm ap} =  0.84'' \times 1''$, equivalent to $\sim$72\% of the
effective radius. 

\begin{figure}
\begin{minipage}{0.5\textwidth}
\begin{center}
\includegraphics[height=6.2cm]{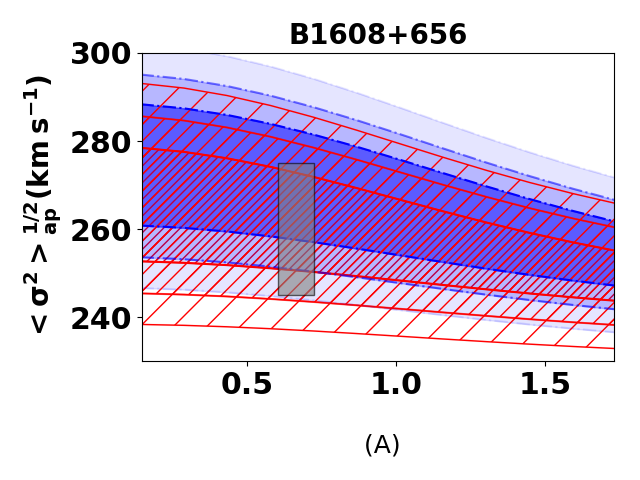}
\end{center}
\end{minipage}
\begin{minipage}{0.5\textwidth}
\begin{center}
\includegraphics[height=6.2cm]{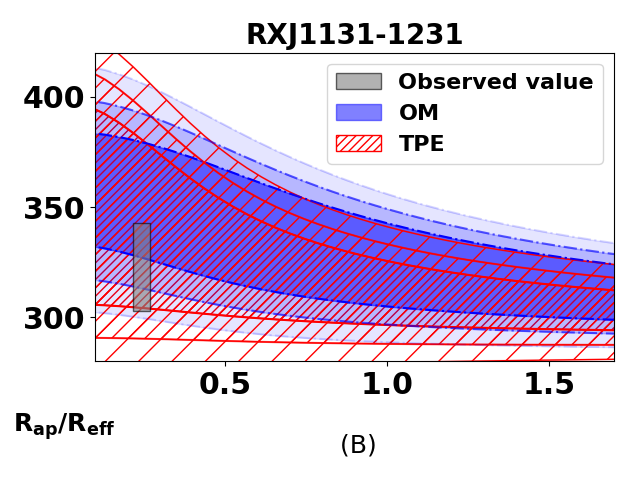}
\end{center}
\end{minipage}
\caption{\textbf{Predicted velocity dispersion as a function of the aperture size, compared to the observed value.} The luminosity-weighted aperture-averaged velocity dispersion
 for the OM (blue shaded) and TPE (red hatched) anisotropy models are shown, with
 normalization factors (mass, angular diameter distance, and the
 Einstein radius) fixed to the best-fitting values. The grey shaded region shows the observed velocity dispersion and the size of the aperture: the vertical position/height of the box shows the measurement of the velocity dispersion ($\langle \sigma^2\rangle^{1/2}_\mathrm{ap}=260\pm15$ \cite{fassnacht/etal:2002}
and $323\pm20$ km\,s$^{-1}$ \cite{sluse/etal:2003} for each lens, respectively), while the horizontal location/width of the box ranges from the shorter half width to the longer half width of the aperture, normalized by the effective radius of the lens galaxy. We vary the slope of
 the mass profiles $\gamma'$ to 1, 2, and 3-$\sigma$ of the posterior probability 
 distribution \cite{suyu/etal:2010, suyu/etal:2013}, shown as the densest to the least dense shaded areas (A)
 B1608+656, and (B) RXJ1131$-$1231. We use flat priors on the anisotropy parameters
 $r_\mathrm{ani}=[0.5,5]$ (OM) and $\beta_\mathrm{in,out}=[-0.6,0.6]$
 (TPE). Our models are compatible with the measurements: the boxes overlap substantially with the 1$\sigma$ regions.
}
\label{fig:sig_apt}
\end{figure}

In \cref{fig:sig_apt}, we illustrate varying sizes of aperture with fixed aspect ratio, 
to show how non-spherically symmetric velocity dispersion change the predicted
aperture-averaged line-of-sight velocity dispersion, in a power-law mass model where the mass density of the galaxy follows $\rho(r) \propto r^{-\gamma'}$, where $\gamma'$ is the slope of the mass profile. We adopt two parameterized models of the
velocity anisotropy, Osipkov-Merritt (OM) \cite{osipkov:1979,
merritt:1985} and a Two-parameter Extension (TPE) of OM
\cite{churazov/etal:2010,agnello/etal:2014} (\cref{sec:vel_ani}). If the aperture had infinite width, the observed velocity dispersion would be the virial
limit where the total kinetic energy of a system can be estimated from its total gravitational potential, thus the relation $GM/R \sim \sigma^2$ holds. In this limit, the uncertainty due to the anisotropy is minimized and
the difference due to the density profile is the only factor determining
the aperture averaged velocity dispersion \cite{agnello/etal:2014}.
The real size of the aperture
is a fraction of the effective radius, so the uncertainty due to the
anisotropy is larger. \cref{fig:sig_apt} shows these compared to the measured velocity dispersion as a function of the aperture size (normalized by the effective radius of the galaxy),
$R_\mathrm{ap}/R_\mathrm{eff}$. 
The TPE model has larger uncertainty than
the OM. 
With the measured uncertainty for the observed velocity dispersions,
the difference between the medians of these two anisotropy models is smaller
than the statistical uncertainties ($\pm 15$ km\,s$^{-1}$ for B1608+656, $\pm
20$ km\,s$^{-1}$ for RXJ1131$-$1231). 
The measured velocity dispersion is itself model dependent e.g. sensitive to a choice of stellar spectral templates \cite{collett/etal:2018}. This leads to a systematic uncertainty in the velocity dispersion measurement which, in turn, affects the angular diameter distance via $D_{\rm d}\propto \sigma^{-2}$. This source of systematic uncertainty is taken into account in the velocity dispersion measurements of B1608+656 and RXJ1131$-$1231.

\begin{figure}
\begin{center}
\subfloat{%
  \includegraphics[clip,height=7cm]{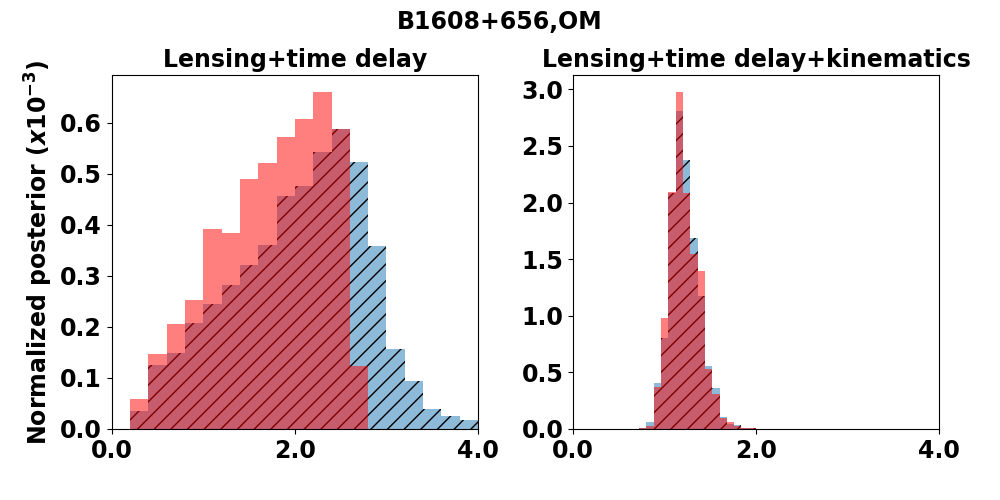}%
}
\\(A)$~~~~~~~~~~~~~~~~~~~~~~~~~~~~~~~~~~~~~~~~~~~~~~~~~~~~~~~~~$(B)
\subfloat{%
  \includegraphics[clip,height=7cm]{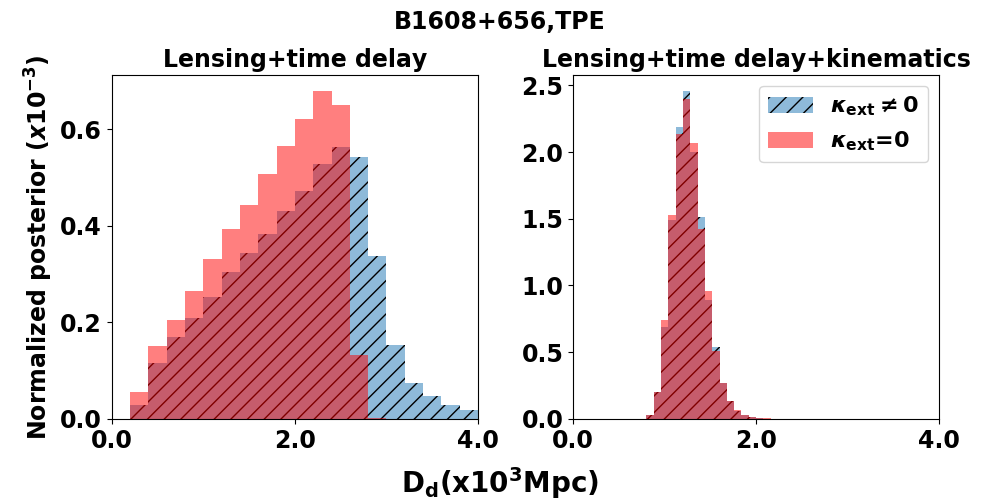}%
}
\\(C)$~~~~~~~~~~~~~~~~~~~~~~~~~~~~~~~~~~~~~~~~~~~~~~~~~~~~~~~~~$(D)

\caption{\textbf{Normalized posterior probability distributions for the angular diameter distance to
 the lens B1608+656.} (A, C) include lensing and time-delay
 information, while (B, D) include additionally the
 kinematics of the lens. The blue hatched distribution shows the results if
 the external convergence distribution is estimated by ray-tracing
 through the Millennium Simulation \cite{hilbert/etal:2009} (\cref{fig:ext_conv}), while the
 red distribution is the result when the external convergence is set to zero. By
 including the kinematic information, the angular diameter distance becomes insensitive to $\kappa_\mathrm{ext}$.}
\label{fig:1608}
\end{center}
\end{figure}
\begin{figure}
\begin{center}
\subfloat{%
\includegraphics[clip,height=7cm]{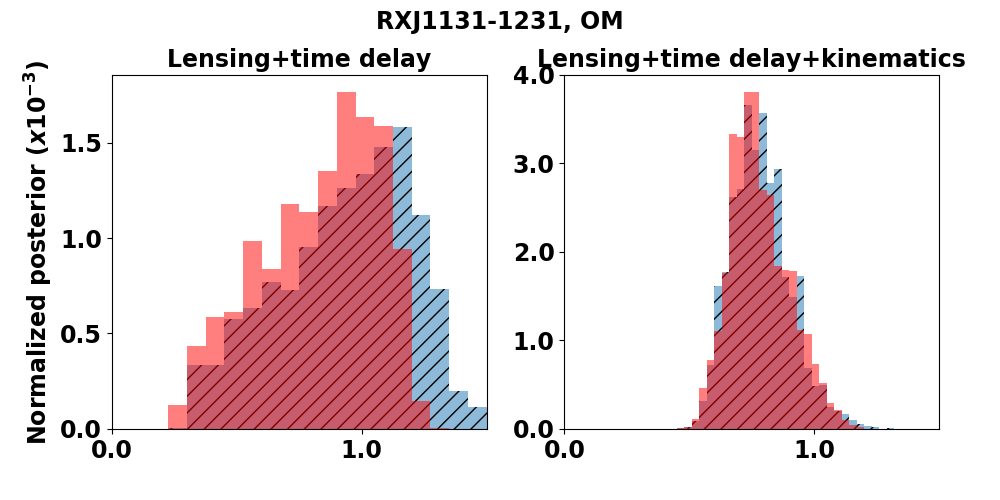}%
}
\\(A)$~~~~~~~~~~~~~~~~~~~~~~~~~~~~~~~~~~~~~~~~~~~~~~~~~~~~~~~~~$(B)
\subfloat{%
  \includegraphics[clip,height=7cm]{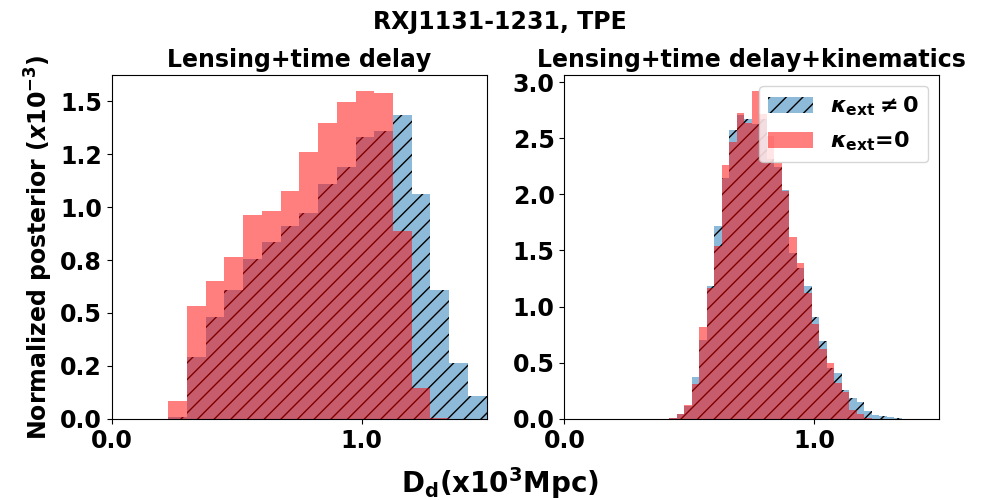}%
}
\\(C)$~~~~~~~~~~~~~~~~~~~~~~~~~~~~~~~~~~~~~~~~~~~~~~~~~~~~~~~~~$(D)
\caption{\textbf{Same as (Fig. 3), but for RXJ1131$-$1231.}  
}
\label{fig:1131}
\end{center}
\end{figure}

\cref{fig:1608} shows the posterior probability
distributions of $D_\mathrm{d}$ of B1608+656 estimated using OM and TPE
anisotropy models (\cref{sec:vel_ani}), without and with the velocity dispersion information. While $\kappa_{\rm ext}$
shifts the posterior probability distribution when the velocity dispersion is not
included, the measurement of $D_{\rm d}$ becomes insensitive to  $\kappa_{\rm ext}$ when it is included
\cite{jee/etal:2015a}. This is because the velocity dispersion
information (with assumed anisotropy) provides additional constraints on
the gravitational potential, which normalizes the angular diameter distance. \cref{fig:1131} shows the same for RXJ1131--1231. 

Our analysis constrains the angular diameter distances to 12--20\% precision per lens. We marginalize over the uncertainties in
anisotropy models by merging two posterior probability distributions of OM and TPE
models (\cref{sec:comb}).
 Our final measurements of the angular diameter distances are 
 $D_\mathrm{d}(z=0.6304)= (1.23 ^{+0.18}_{-0.15})\times 10^3~{\rm Mpc}$ for B1608+656, and $D_\mathrm{d}(z=0.295)= (8.1 ^{+1.6}_{-1.2} )\times 10^2~{\rm Mpc}$ for RXJ1131$-$1231.

\begin{figure}
\begin{center}
\includegraphics[height=12cm]{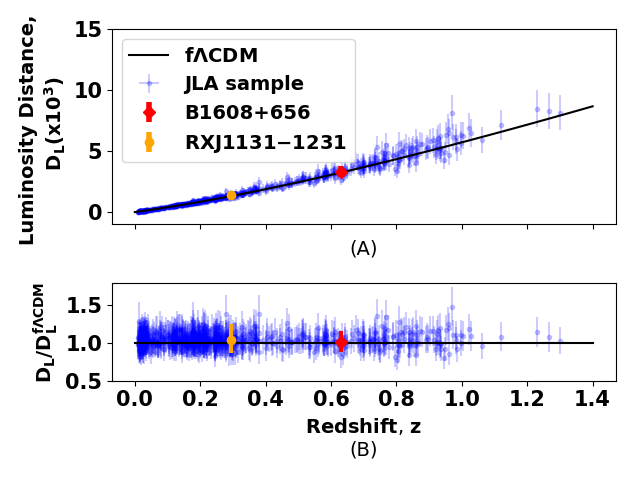}
\caption{\textbf{Derived Hubble diagram (A) and its residuals (B).} The blue points with errorbars are 740 SNe from JLA \cite{betoule/etal:2014}, normalized by our two
 lensing distances, shown as orange and red point. The solid line is the best-fitting flat $\Lambda$CDM model.}
 \label{fig:hubble}
\end{center}
\end{figure}

We apply these distances as anchors to the 740 SNe in the Joint Light-curve Analysis (JLA, \cite{betoule/etal:2014})
dataset, allowing us to constrain
$H_0$ and the SNe nuisance parameters (\cref{sec:sne}) simultaneously. We use the MontePython code
\cite{Audren/etal:2012} to perform a Markov Chain Monte Carlo
analysis. \cref{fig:hubble} shows the resulting Hubble diagram, i.e. the absolute luminosity
distances $D_\mathrm{L}=(1+z)^2D_\mathrm{d}$ as a function of redshifts for a flat
$\Lambda$CDM model. 

\begin{figure}
\begin{center}
\includegraphics[height=9cm]{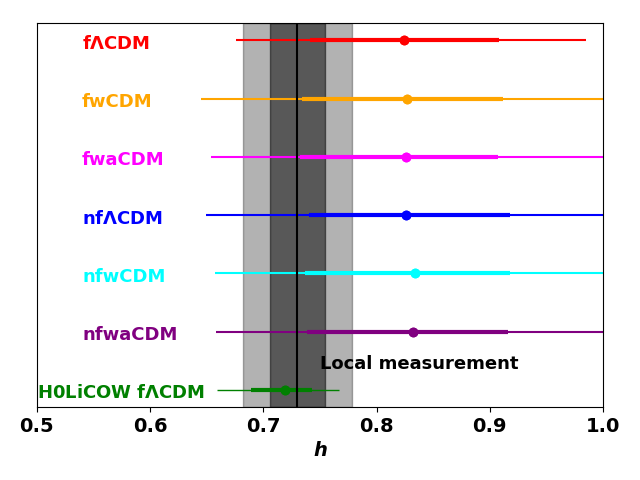}
\caption{\textbf{Constraints on the Hubble constant for six cosmological models.} The gray shaded area is the constraint from the local distance
 ladder \cite{riess/etal:2016}, while the green line is from three time-delay distances measured by the H0LiCOW collaboration \cite{bonvin/etal:2016}. The thick and thin solid lines denote the 68\% and 95\% Confidence Levels of the joint fit to the SNe and the $D_{\rm d}$ data. We emphasize that $D_\mathrm{d}$ and $D_{\Delta t}$ determined from the same lens are not correlated strongly because the uncertainty in the former is dominated by the kinematics and the latter by $\kappa_{\rm ext}$ in the case of our two lenses. Therefore, the corresponding constraints on $H_0$ are nearly independent.}
 \label{fig:H_comp}
\end{center}
\end{figure}

\cref{fig:H_comp} shows the inferred values of $H_0$ assuming various
cosmological models: $\Lambda$CDM with flat spatial geometry (f$\Lambda$CDM) and
non-flat spatial geometry (nf$\Lambda$CDM); a dynamical dark energy model with flat spatial 
geometry (f$w$CDM) and non-flat spatial geometry (nf$w$CDM), where $w$ is the dark energy equation of state that characterizes the time evolution of dark energy density, and $w$ is a parameter in these models; and a dynamical
dark energy model with a time-varying equation of state ($w(z) = w_0 +
zw_a/(1+z)$) with flat spatial geometry (f$w_a$CDM) and non-flat spatial geometry (nf$w_a$CDM), where $w_0$ and $w_a$ are parameters. 
By construction the inverse distance ladder method
is insensitive to the assumed cosmological models, which is reflected by the consistent values in \cref{fig:H_comp}. Therefore, we
adopt the value for flat $\Lambda$CDM, 
$H_0=82.4^{+8.4}_{-8.3}~{\rm km\,s^{-1}\,Mpc^{-1}}$ (68\% CL) as our fiducial result.
We examine and marginalize over uncertainties in the kinematics and 
mass profiles of the lens galaxies (\cref{sec:vel_ani}).
All values of $H_0$ we obtain are consistent with
$H_0$ from the distance ladder method \cite{riess/etal:2016} and
from the time-delay distances \cite{sluse/etal:2016, rusu/etal:2017,wong/etal:2017,bonvin/etal:2016}. 
It is also consistent with, but more precise than, $H_0$ from
the standard siren method \cite{schutz:1986,abbott/etal:2017}.

While our measurement has a larger uncertainty than other direct
methods, this is dominated by statistical uncertainty because we use only two lenses to normalize the SNe
distances. The precision in our $H_0$ measurement is presently limited by the number of strong lens systems with measured time delays and ancillary data. 
Systematic errors, although subdominant, are mainly due to the determination of the velocity structure of the lenses. The single aperture-averaged kinematic measurement and modeling present the main systematic error, which can be overcome by e.g., from
spatially resolved kinematic data.

 

\clearpage

\bibliography{scibib}

\bibliographystyle{Science}
\clearpage

\section*{Acknowledgements}
We thank the H0LiCOW and COSMOGRAIL collaborations for providing access to the observational data. We thank Simon Birrer for many comments which helped making the analysis more thorough and complete. We thank Alessandro Sonnenfeld and Matthew Auger for providing their velocity dispersion code. We also thank Ak\i n Y\i ld\i r\i m for assistance with the Jeans Anisotropic Model code. 
\section*{Funding}
SHS thanks the Max Planck Society for support through the Max Planck Research Group. CDF acknowledges support from the National Science Foundation grant AST-1312329. SH acknowledges support by the DFG cluster of excellence `Origin and Structure of the Universe'
(www.universe-cluster.de). LVEK is supported in part through an NWO-VICI career grant (project number 639.043.308). 
\section*{Author contributions}
All authors participated in the design of the experiment, the interpretation of the data, and the writing of the manuscript. 
IJ performed the dynamics modeling and statistical analysis for inferring $D_\mathrm{d}$, the inverse distance ladder analysis combining the JLA SNe with the lensing distances, drafting and revising the manuscript. 
SHS performed the lens mass modeling using the lensing and time-delay data, Bayesian statistical modelling combining the kinematic data set with the lensing, time-delay and external convergence data sets for inferring $D_\mathrm{d}$, and assisted in the dynamic modeling of the lens.
EK conceived and oversaw this research project. 
CDF was the principal investigator of the observing programs that obtained the Hubble Space Telescope (HST) imaging of B1608+656 and the Keck spectroscopy for both lens systems and, led the acquisition and initial reduction of those data. SH obtained and analyzed simulation data for the external convergence estimation. 
LVEK contributed to the HST and kinematics data sets and their analyses, the lensing and kinematic modelling of the presented lenses, the interpretation 
of the results and comments on the manuscript.
\section*{Competing interest}
The authors have no competing interests.
\section*{Data and materials availability}
The Hubble imaging data is publicly available at the Hubble Legacy Archive\\ (https://archive.stsci.edu/hst/search.php), under Proposal IDs 10158 (PI: Fassnacht), 7422 (PI: Readhead) and 9744 (PI: Kochanek). All the other data, e.g. the velocity dispersion, are available in the manuscript, supplementary material or the references therein.  Our scripts and input files for our analysis are publicly available at https://github.com/jee1213/ScienceInvDistLadder. Our output cosmological parameters are in Tables S1 and S2, with more details available at the same weblink (https://github.com/jee1213/ScienceInvDistLadder).  To compute the velocity dispersions, we have used codes developed by Matthew Auger and Alessandro Sonnenfeld that are available at https://github.com/astrosonnen/spherical\_jeans. 

\clearpage

\setcounter{page}{1}
\include{Material_method_rev2}

\clearpage

\end{document}

%% file: Material_method_rev2.tex









\section*{Materials and Methods}
\renewcommand{\thefigure}{S\arabic{figure}}
\setcounter{figure}{0}
\renewcommand{\thetable}{S\arabic{table}}
\renewcommand{\theequation}{S\arabic{equation}}
\renewcommand{\thesection}{S\arabic{section}}
\setcounter{table}{0}
\section{Anisotropic Velocity Dispersion Models}
\label{sec:vel_ani}
Our method relies on the ability of the measured stellar velocity
dispersion to constrain the gravitational potential of the lens
galaxy. However, the gravitational potential of a system is related only
to the radial component of the velocity dispersion, $\sigma_\mathrm{r}^2$, via the spherical Jeans equation:
\begin{equation}
\frac{1}{\rho_*}\frac{d(\rho_*\sigma_\mathrm{r}^2)}{dr} +2 \beta_\mathrm{ani}(r)\frac{\sigma_\mathrm{r}^2}{r} = -\frac{GM(r)}{r^2}.
\label{eq:jeans}
\end{equation}
Here, $r$ is the 3-dimensional radius, $\rho_*$ is the stellar density, $\beta_\mathrm{ani}$ is the anisotropy parameter, and $M(r)$ is the total mass enclosed within the radius $r$.
However, the observable quantity is the velocity dispersion
projected along the line-of-sight to the observer. This is also weighted
by the luminosity as it is traced by luminous components; thus, the
observable quantity can be written as
\begin{equation}
\sigma_\mathrm{p}^2(R) = 2\int^{\infty}_R \left[1-\beta_\mathrm{ani}(r)\frac{R^2}{r^2}\right]\frac{\rho_*(r)\sigma_\mathrm{r}^2(r) r dr}{\sqrt{r^2-R^2}},
\end{equation}
where $R$ is the projected radius. When $\beta_\mathrm{ani}(r)$ is modeled, the above relation
connects the observed $\sigma_\mathrm{p}$ to $\sigma_\mathrm{r}$.

We test two parametric anisotropy models: Osipkov-Merrit
(OM) and its two-parameter extension (TPE). The posterior probability distributions
of the angular diameter distance show a slight dependence on the anisotropic velocity dispersion model. Anisotropy parameters relate the radial and tangential components of the velocity dispersion ellipsoid ($\sigma_\mathrm{r}$ and $\sigma_\mathrm{T}$, respectively). The general form of the anisotropy parameter is 
\begin{equation}
\beta_\mathrm{ani}(r) = \frac{\beta_\mathrm{in}r_\mathrm{a}^2+\beta_\mathrm{out}r^2}{r_\mathrm{a}^2+r^2}\equiv1-\frac{\sigma_\mathrm{T}^2(r)}{\sigma_\mathrm{r}^2(r)},
\label{eq:tpe}
\end{equation}
where $r_\mathrm{a}$ is an anisotropy radius. We see that
$\beta_\mathrm{ani}\rightarrow \beta_\mathrm{in}$ toward the center, and
$\beta_\mathrm{ani}\rightarrow \beta_\mathrm{out}$ toward the outskirts
of the galaxy. The OM model takes
$\beta_\mathrm{in}=0$  and $\beta_\mathrm{out}=1$, so as $r \rightarrow
0$ the velocity dispersion becomes isotropic, while at the outskirts the
velocity dispersion becomes radial. The OM model produces 
positive-valued phase space distribution function by construction, thus any
parameter choice is physically meaningful. However, the OM model does not
cover all possible anisotropy structures. For the TPE model, we adopt
flat priors on both $\beta_\mathrm{in}$ and $\beta_\mathrm{out}$ in the range
$[-0.6,0.6]$. These choices are conservative relative to the posterior
distributions of the parameters in kinematics modelling \cite{agnello/etal:2014} of the nearby massive elliptical galaxy
M87 using globular clusters. The TPE model probes a larger variety of anisotropy
structure in comparison to the OM model, but is not guaranteed to have
a positive phase space distribution function at all radii.

We solve the spherical Jeans equation with this anisotropy
parameter. Assuming the stellar density follows a Hernquist profile
\cite{hernquist:1990}, we calculate the aperture-averaged
luminosity-weighted projected velocity dispersion. We use this quantity
to infer the gravitational potential of the lens. Another stellar
density model, the Jaffe profile \cite{jaffe:1983}, was tested with a
power-law total mass density in \cite{jee/etal:2015a}. We find that the luminosity-weighted velocity dispersion of the Jaffe profile depends less on the
velocity dispersion anisotropy than that of the Hernquist profile, thus
our choice of the Hernquist profile as the 
stellar density is more conservative.

For B1608+656, the $1\sigma$ uncertainties on the measured angular
diameter distance are similar for both models: for the OM profile it is
13\%, and for the TPE model it is 14\%. However, for RXJ1131$-$1231, it
is 14\% for the OM profile while the uncertainty increases to 18\% for
the TPE model. The current data have only the aperture-averaged
velocity dispersion available, so do not favor either anisotropy model
over the other. Thus, we combine the posterior distance
measurements from each anisotropy model for the cosmological
implication analysis (see below).
\section{Calculating Velocity Dispersion}
We used kernel method \cite{mamon/lokas:2004} to calculate the velocity dispersion. This requires first solving the Jeans equation (\cref{eq:jeans}), then projecting it along the line-of-sight by integrating the solution to obtain the luminosity-weighted projected velocity dispersion. This two-step process can be combined into a single kernel integration for various anisotropy models \cite{mamon/lokas:2004}. In this section, we derive the kernel for the TPE model and discuss the limitation of this method when applied to the TPE model.

The general form of Jeans equation can be rewritten as,
\begin{equation}
    l(r)\sigma^2_\mathrm{r}(r) = \frac{1}{f(r)}\int_{r}^{\infty} f(s)l(s)\frac{GM(r)}{s^2} ds
\end{equation}
where $f(r)$ satisfies 
\begin{equation}
    \frac{d \ln f(r,r_a)}{d \ln r} = 2\beta_\mathrm{ani}(r, r_a)
    \label{eq:f}
\end{equation}
[\cite{mamon/lokas:2004}, their appendix A] and $l(r)$ is the light distribution. From Equations \ref{eq:f} \& \ref{eq:tpe}, $f(r,r_a)$ and $df(r,r_a)/dr$ in the TPE model become
\begin{equation}
    f(r,r_a) = r^{2 \beta_\mathrm{in}}(r^2+r_\mathrm{a}^2)^{\beta_\mathrm{out}-\beta_\mathrm{in}}
    \label{eq:fr}
\end{equation}
and
\begin{equation}
    \frac{df(r,r_a)}{dr} = 2f(r,r_a)\left[\frac{b_\mathrm{in}}{r}+\frac{r(\beta_\mathrm{out}-\beta_\mathrm{in})}{r^2+r_a^2}\right].
    \label{eq:dfdr}
\end{equation}
Equations \ref{eq:fr} \& \ref{eq:dfdr} can then be used to calculate the kernel $K$ [\cite{mamon/lokas:2004}, their appendix A]:
\begin{equation}
    K(u,u_a) = \frac{2}{u}f(u,u_a)\int_{1}^{u}du' \frac{u'-\frac{1}{2f(u',u_a)}\frac{df(u',u_a)}{du'}}{f(u',u_a)\sqrt{u'^2-1}},
    \label{eq:kernel}
\end{equation}
with the change of variables $u \equiv r/R $ and $u_a \equiv r_a/R$. The line-of-sight velocity dispersion can be expressed as
\begin{equation}
    I(R)\sigma^2_{los}(R) = 2G\int^{\infty}_{R}K\left(\frac{r}{R},\frac{r_a}{R}\right)l(r)M(r)\frac{dr}{r}.
    \label{eq:Isig}
\end{equation}
For the anisotropic models \cite{mamon/lokas:2004}, \cref{eq:kernel} can be approximated analytically and therefore computing \cref{eq:Isig} can be made substantially faster. However, for the case of the TPE model, the analytic integration in \cref{eq:kernel} involves Appell's hypergeometric function of the first kind \cite{Appell:1925}, $F_1(\alpha;\beta,\beta';\gamma;x,y)$, where $x$ and $y$ are $-u^2/u_a^2$ and $u^2$. The series definition of this hypergeometric function is convergent only in the region $|x| < 1$ and $|y| < 1$.
Using analytic continuation, the function can be evaluated outside the region (e.g. \cite{colavecchia/etal:2001}), but its implementation involves Gauss hypergeometric functions and is not necessarily faster to evaluate computationally. Thus, we have numerically estimated \cref{eq:Isig} with double integral.

In computing the velocity dispersion from spherical Jeans equation, we assume a spherical mass distribution for the lens galaxy. Nonetheless, the lens galaxies in our lens systems appear elliptical and are described by elliptical surface mass density distributions in the lens mass modeling. To test the impact of our spherical Jeans modeling assumption on our $D_{\rm d}$ measurement, we use the dynamical modeling machinery of \cite{yildirim/etal:2019}. We model both the lensing and aperture-averaged velocity dispersion data self-consistently with an axisymmetric power-law mass model (instead of spherical mass model for the kinematics) using Jeans Anisotropic Models (JAM) \cite{cappellari:2008, barnabe/etal:2012, vandeven/etal:2010}. As the ellipticity of the lens galaxy in RXJ1131$-$1231 is small (with axis ratio $b/a\sim0.8$; \cite{suyu/etal:2012}), the velocity dispersion changes only by a few km\,s$^{-1}$ in comparison to the case where we force the system to be spherical in the dynamical modeling ($b/a=1$). This is within the measurement uncertainty of 20 km\,s$^{-1}$.  In the case of B1608+656, the system is more elliptical ($b/a\sim0.6$; \cite{koopmans/etal:2003}), and the change can potentially be up to the 1 $\sigma$ level based on our estimate from RXJ1131$-$1231. However this effect is still within the measurement uncertainty, as the kinematics estimate we are using is from an aperture-averaged measurement. Therefore, for our $D_{\rm d}$ measurements based on single aperture-averaged velocity dispersion, sensitivity to non-sphericity is weak with limited impact on $D_{\rm d}$ given the present uncertainties.

\section{Baryon-Dark matter Composite Lens Mass Model}
\label{sec:composite}
\begin{figure}
    \centering
    \includegraphics[clip,height=9cm]{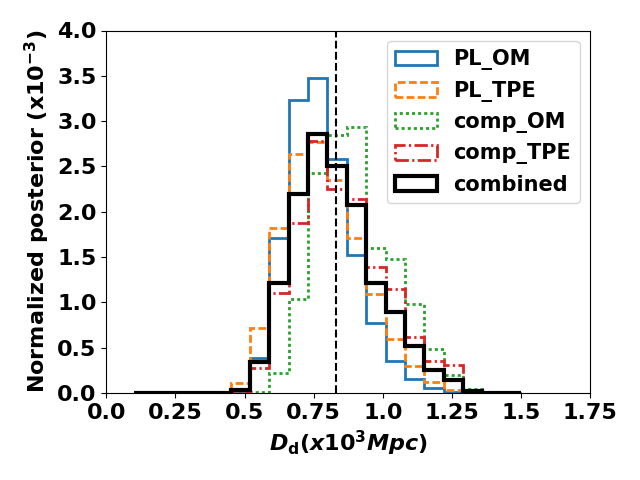}
    \caption{\textbf{Combined $D_{\rm d}$ distribution for RXJ1131$-$1231.} $D_\mathrm{d}$ measurements comparison between two mass models, power-law (PL) and composite (comp) and two anisotropy models, OM and TPE. The combined distribution yields $D_\mathrm{d} = 810 ^{+160}_{-130} $ Mpc.}
    \label{fig:compPL}
\end{figure}

We assumed a power-law profile for our fiducial mass model. Assuming a functional form of the mass profile may artificially break the lens mass model degeneracy due to source-position transformation (SPT) \cite{schneider/sluse:2013a}. For example, a baryon-dark matter composite model can reproduce almost the same image position and surface brightness while still giving different time delay, thus yields different $H_0$ constraints \cite{schneider/sluse:2013a}. However, quantitative calcultions \cite{wertz/etal:2017} have shown that the impact of SPT on measured time-delay distance is a few percent. To test the robustness of $D_\mathrm{d}$ estimation under SPT, we used a composite model \cite{suyu/etal:2013} for RXJ1131$-$1231, and quantify the effect of different lens mass profile on the measured angular diameter distance. In this model, the baryon and dark matter are parameterized separately: the baryonic component follows a Sersic profile \cite{sersic:1963} under the assumption that the mass-to-light ratio is  spatially constant. The dark matter component follows a Navarro, Frank and White (NFW) profile \cite{navarro/frank/white:1995}, where its concentration is parameterized by the scale radius $r_s$. The circularly averaged convergence profile of this composite model is similar to that of the power-law profile at radii $0.5''<r<3''$, where the lensing arcs are observed \cite{suyu/etal:2013}. However,  by combining the lensing, time-delay and kinematics data, both models yield very similar $D_{\Delta t}$. To test if this also holds for the angular diameter distance, we compare $D_\mathrm{d}$ constraints from power-law and  composite models in \cref{fig:compPL} for both the OM and TPE anisotropy velocity dispersion models.
While the agreement between $D_\mathrm{d}$ from different mass models is not as good as that of $D_{\Delta t}$, the result shows that $D_\mathrm{d}$ measurement from four different models agree well within 1-$\sigma$. The combined distribution yields $D_\mathrm{d} = (8.1^{+1.6}_{-1.3}) \times 10^2$ Mpc. We adopt this as the $D_\mathrm{d}$ measurement and its uncertainty for RXJ1131$-$1231, and use it for the cosmological constraints analysis.
 Pixelated lens potential corrections on power-law models of B1608+656 are small ($\sim$2\%) \cite{suyu/etal:2009} indicating that power-law models provide a sufficient description of the lens system. This validates our use of the power-law models for this system.

\section{Truncated Lens Mass Distribution}
A conventional composite lens mass model with an infinitely extended NFW dark matter halo has been tested against a more realistic halo with a truncation radius \cite{Munoz:2017}. At small radii, typically a few kpc from the center of the lens galaxy where strong lensing arcs are observed, the impact of the outer mass truncation on lensing observable has an effect almost identical to that of a mass-sheet transformation (MST) to the lens mass model  \cite{Munoz:2017}. The halo truncation impacts the time-delay distance constraints in the same way as MST, resulting in a biased estimate of $D_{\Delta t}$, but the bias is $\leq 1\%$ \cite{Munoz:2017}. However, $D_\mathrm{d}$ is unaffected by the uniform mass sheet \cite{jee/etal:2015a}, and thus the halo truncation does not affect the measured $D_\mathrm{d}$.

As $r_s$ changes, the lens model parameters also vary. We thus constrain the lens model parameters, Einstein radius $\theta_{\rm E}$, scale radius $r_s$, baryon $M/L$ ratio and $D_{\Delta t}$, using a lens modelling code {\sc Glee} \cite{suyu/halkola:2010,glee:suyu/etal:2012} given different priors on scale radii from the imaging data. The prior on scale radius is either: (1) a flat prior in range $8''<r_s<40''$, or (2) a Gaussian prior as $r_s = G(\mu=18''.6, \sigma=2''.6)$, where $\mu$ is the mean and $\sigma$ is the standard deviation of the Gaussian. Then, assuming that the posterior of these lens model parameters follow a Gaussian distribution, we calculate the covariance matrix of the parameters constrained from the lens model. We use the statistical technique of importance sampling \cite{lewis/bridle:2002, suyu/etal:2010} (\cref{sec:ImpSamp}) to obtain the angular diameter distance while sampling the lens model parameters from the covariance matrix and see how different priors on $r_s$ change the distance. We test this for the composite mass model, and two anisotropic velocity dispersion models that we previously used, OM and TPE.

\begin{figure}
    \centering
    \includegraphics[clip,height=7cm]{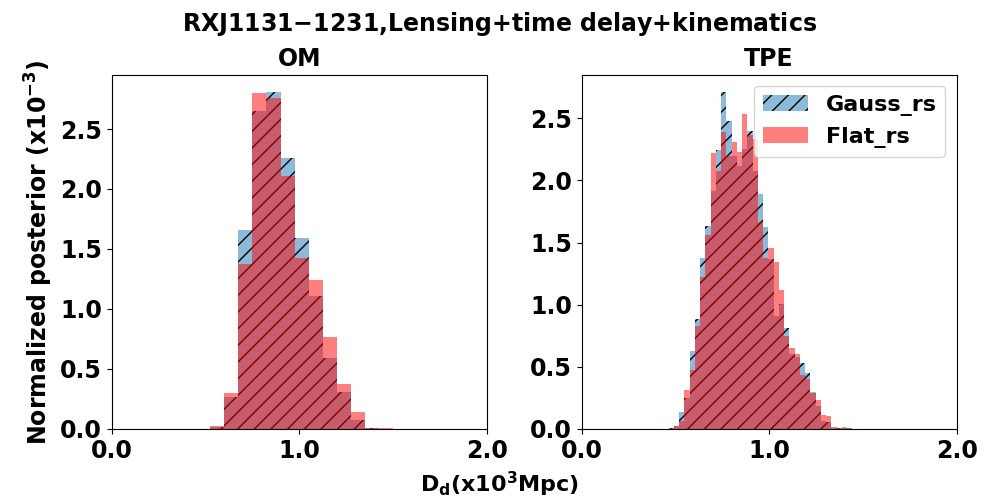}
    \\(A)$~~~~~~~~~~~~~~~~~~~~~~~~~~~~~~~~~~~~~~~~~~~~~~~~~~~~~~~~~~~$(B)
    \caption{\textbf{Effect of dark matter halo truncation on $D_{\rm d}$.} Comparison between $D_\mathrm{d}$ distributions for a composite lens mass model with a Gaussian prior (blue hatched) and a flat prior (red) on $r_s$ for OM (A) and TPE (B) anisotropy models. For both OM and TPE models, the red and the blue distributions are almost identical, showing that the effect of dark matter halo truncation on $D_\mathrm{d}$ constraints is negligible.}
    \label{fig:Dd_rs}
\end{figure}

The results are shown in \cref{fig:Dd_rs}. The distributions are
almost identical, indicating that the halo truncation does not affect the $D_\mathrm{d}$ constraints. Thus, we use $D_\mathrm{d}$ constrained using a Gaussian $r_s$ prior in Section \ref{sec:composite} when calculating the final $D_\mathrm{d}$ for RXJ1131$-$1231.

\section{Importance Sampling the Lensing Likelihood with Dynamical Models}
\label{sec:ImpSamp}
To constrain the angular diameter distance using observational data, we
use the posterior probability density function (PDF) of the lens model
parameters obtained 
using the lensing and time-delay data of B1608+656 \cite{suyu/etal:2010}
and RXJ1131$-$1231 \cite{suyu/etal:2013}.
For the kinematics of the lens, we importance sample
\cite{lewis/bridle:2002, suyu/etal:2010} the posterior PDF of the lens
mass model parameters with the likelihood of the kinematics data. 
Therefore, we take the previously-published lensing and time-delay mass models \cite{suyu/etal:2013,suyu/etal:2010}, and importance sample these results using the velocity dispersion measurements and our kinematic models.
The model velocity dispersion is luminosity-weighted
and aperture-averaged to be compared with the observation. In this
section we briefly summarize the data, models, and the marginalization
process.

Time-delay lensing cosmography requires multiple types of observations
per lens, each providing different information about the lens. First,
the imaging data provide the surface brightness distribution of the lens
and the lensed source in a pixelated form, which enables the
reconstruction of the Fermat potential via modeling of the lens mass
distribution. For both lenses, we use Hubble Space Telescope (HST) Advanced Camera for Surveys (ACS)
data. The imaging data are denoted as
$\boldsymbol{d_\mathrm{ACS}}$. These data include the lens light and the
lensed source light. The latter is separated into an extended host galaxy surface
brightness and a point-like Active Galactic Nucleus (AGN) light for each AGN image. These
components are blurred by the point spread function, which are
modeled using stars in the field of view. Long-term monitoring of the
variability of individual images provides the time delay between image
pairs, $\boldsymbol{\Delta t}$. For B1608+656, the time delay is
taken from previous publications \cite{fassnacht/etal:1996, fassnacht/etal:2002}. For
RXJ1131$-$1231, we use the time-delay measurement published in \cite{tewes/etal:2013}, which is obtained by monitoring the lens system in optical wavelengths. The lens and source redshifts are, 
respectively,  $z_\mathrm{d} = 0.6304$ \cite{myers/etal:1995} and $z_\mathrm{s} = 1.394$ \cite{fassnacht/etal:1996}
 for B1608+656; $z_\mathrm{d} = 0.295$ \cite{sluse/etal:2003, sluse/etal:2007} and $z_\mathrm{s} = 0.654$ \cite{sluse/etal:2007}
for RXJ1131$-$1231. Further spectroscopic data for the lenses are obtained from the Low-Resolution Imaging
Spectrometer (LRIS) on the Keck I telescope, providing the luminosity-weighted,
aperture-averaged line-of-sight velocity dispersion of the lens,
$\langle\sigma^2_{ap}\rangle^{1/2}$. For convenience, we denote this
simply as $\sigma$ in this section. We adopt the published
velocity dispersion \cite{suyu/etal:2013}. In measuring the time-delay distance, the lens
environment, $\boldsymbol{d_\mathrm{env}}$, should be modeled to break
the mass-sheet degeneracy \cite{falco/etal:1985, suyu/etal:2010,
schneider/sluse:2013a, suyu/etal:2013}. This provides information on the
external convergence $\kappa_\mathrm{ext}$.

The lens mass distribution is modeled as an elliptical power-law profile
with external shear $\gamma_{\rm ext}$. The mass density
 profile slope ($\gamma'$, with the 3D mass density $\rho(r)\propto
 r^{-\gamma'}$) is constrained by the lensing analysis in range
 $\gamma'=2.08\pm0.03$ for B1608+656 and $\gamma'=1.95^{+0.05}_{-0.04}$ for
 RXJ1131$-$1231. The lens light is modeled as
 elliptical S\'ersic profiles, and the lensed source light as an
 independent point-source AGN per image plus a non-parametric extended
 source surface brightness estimated on a regular grid at the source
 plane. Once the lens mass distribution is constrained from the image, the time delay between image pairs given the lens mass model can be calculated. Thus, time delay pairs provide some constraints on the lens model as well as constraining the time-delay distance \cite{suyu/etal:2010, suyu/etal:2013}. We model the velocity dispersion using two parametric anisotropy models as introduced above. The external convergence is estimated based on the relative number count of galaxies neighboring the lens with respect to the field galaxies, and calibrated using the Millennium Simulation \cite{hilbert/etal:2009}.

Under the Bayesian framework, we find the posterior PDF of the model
parameter set $\boldsymbol{\xi}$ given the choice of physically
motivated models for each component,
$P(\boldsymbol{\xi}|\boldsymbol{d_\mathrm{ACS}},\sigma,\boldsymbol{d_\mathrm{env}},\boldsymbol{\Delta
t})$. The model parameters are $\boldsymbol{\xi} = \{\boldsymbol{D},
\gamma', \theta_\mathrm{E}, \gamma_\mathrm{ext}, \boldsymbol{\eta},
\boldsymbol{\beta_\mathrm{ani}}, \kappa_\mathrm{ext}\}$, where
$\boldsymbol{D} = \{D_\mathrm{d},D_\mathrm{ds}/D_\mathrm{s}\}$, and
$\boldsymbol{\eta}$ is a vector containing the rest of the lens model
parameters. Two differences between the parameters we use here and the
previous analysis \cite{suyu/etal:2010, suyu/etal:2013} are: i)
Instead of calculating the posterior PDF of cosmological parameters, we
choose to directly calculate the posterior PDF of $D_\mathrm{d}$ and
$D_\mathrm{ds}/D_\mathrm{s}$. This is because we want to constrain the
angular diameter distance alone. 
ii) We test two velocity dispersion anisotropy models, thus
instead of using $r_\mathrm{ani}$, we use
$\boldsymbol{\beta_\mathrm{ani}}$.

\begin{figure}[t]
\centering
\includegraphics[width=16cm]{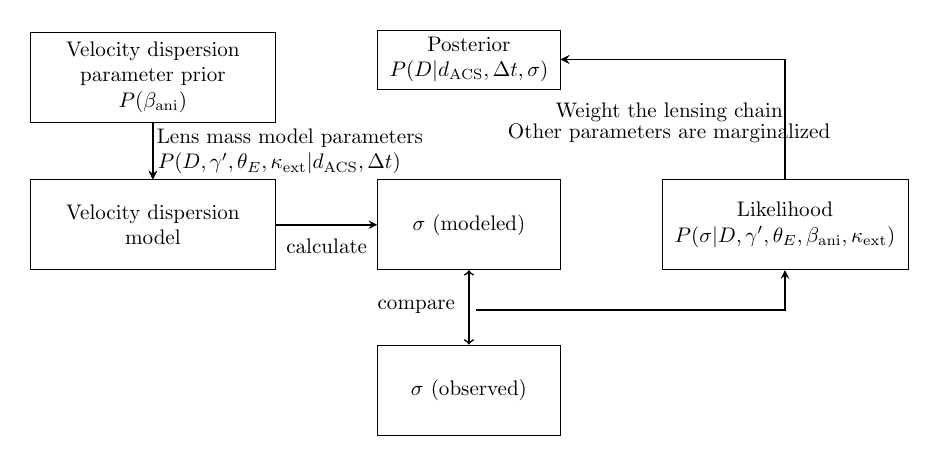}
\caption{\label{fig:schem} \textbf{Schematic diagram of importance sampling.} The dynamical model is separately calculated and incorporated into the lensing likelihood, to constrain the posterior PDF of the distances.}
\end{figure}

Bayes' theorem states
\begin{equation}
P(\boldsymbol{\xi}|\boldsymbol{d}) = \frac{P(\boldsymbol{\xi})~P(\boldsymbol{d}|\boldsymbol{\xi})}{E(\boldsymbol{d})},
\label{eq:bayse}
\end{equation}
where $\boldsymbol{\xi}$ is the parameter vector, $\boldsymbol{d}$ is
 the data vector, $P(\boldsymbol{d}|\boldsymbol{\xi})$ is the likelihood
 of the data $\boldsymbol{d}$ given the parameters,
 $P(\boldsymbol{\xi})$ is a prior distribution,
 and $E(\boldsymbol{d})$ is the model evidence. The evidence can be used
 to compare different models, but for our purpose it can be treated as a
 proportionality constant that normalizes the posterior PDF. We thus obtain
\begin{equation}
P(\boldsymbol{\xi}|\boldsymbol{d_\mathrm{ACS}},\sigma,\boldsymbol{d_\mathrm{env}},\boldsymbol{\Delta t}) \propto P(\boldsymbol{d_\mathrm{ACS}},\sigma,\boldsymbol{d_\mathrm{env}},\boldsymbol{\Delta t}|\boldsymbol{\xi})P(\boldsymbol{\xi}).
\label{eq:lensing_lkl}
\end{equation}
As all observations are independent, the likelihood in \cref{eq:lensing_lkl} can be separated as 
\begin{equation}
P(\boldsymbol{d_\mathrm{ACS}},\sigma,\boldsymbol{d_\mathrm{env}},\boldsymbol{\Delta t}|\boldsymbol{\xi}) = P(\boldsymbol{d_\mathrm{ACS}}|\boldsymbol{\xi})P(\sigma|\boldsymbol{\xi})P(\boldsymbol{d_\mathrm{env}}|\boldsymbol{\xi})P(\boldsymbol{\Delta t}|\boldsymbol{\xi}).
\label{eq:data_indep}
\end{equation}
Writing the data dependency of the parameters explicitly, we obtain
\begin{equation}
\begin{split}
P(\boldsymbol{d_\mathrm{ACS}},\sigma,\boldsymbol{d_\mathrm{env}},\boldsymbol{\Delta t}|\boldsymbol{\xi}) & = P(\boldsymbol{d_\mathrm{ACS}}|\boldsymbol{D},\gamma',\theta_\mathrm{E},\gamma_\mathrm{ext},\boldsymbol{\eta},\kappa_\mathrm{ext})P(\sigma|\boldsymbol{D},\gamma',\theta_\mathrm{E},\boldsymbol{\beta_\mathrm{ani}},\kappa_\mathrm{ext})\\
 & \times P(\boldsymbol{d_\mathrm{env}}|\gamma_\mathrm{ext},\kappa_\mathrm{ext})
P(\boldsymbol{\Delta t}|\boldsymbol{D},\gamma',\theta_\mathrm{E},\gamma_\mathrm{ext},\boldsymbol{\eta},\kappa_\mathrm{ext}).
\end{split}
\end{equation}
To obtain the posterior PDF of $\boldsymbol{D}$, we marginalize the full
posterior
$P(\boldsymbol{\xi}|\boldsymbol{d_\mathrm{ACS}},\sigma,\boldsymbol{d_\mathrm{env}},\boldsymbol{\Delta
t})$ over the other model parameters,
\begin{equation}
P(\boldsymbol{D}|\boldsymbol{d_\mathrm{ACS}},\sigma,\boldsymbol{d_\mathrm{env}},\boldsymbol{\Delta t}) = \int d\gamma'~ d\theta_\mathrm{E} ~d\gamma_\mathrm{ext}~ d\boldsymbol{\eta}~ d\boldsymbol{\beta_\mathrm{ani}}~ d\kappa_\mathrm{ext} P(\boldsymbol{\xi}|\boldsymbol{d_\mathrm{ACS}},\sigma,\boldsymbol{d_\mathrm{env}},\boldsymbol{\Delta t}),
\end{equation}
by importance sampling. Importance sampling states that the expectation value of a function, $f(x)$, where $x$ follows the PDF $P_1(x)$ can be calculated in the following way:
\begin{equation}
\langle f(x) \rangle_{P_1} = \int P_1(x) f(x) dx = \int \frac{P_1(x)}{P_2(x)}P_2(x)f(x)dx = \langle \frac{P_1(x)}{P_2(x)} f(x) \rangle_{P_2} ,
\end{equation}
where only $P_2(x)$ is available and $P_1(x)$ is not. In combination
with Bayes' theorem, importance sampling allows us to separate the data
in the posterior distribution. In our case, $P_1 =
P(\boldsymbol{D},\gamma',\theta_\mathrm{E},\gamma_\mathrm{ext},\boldsymbol{\eta},\boldsymbol{\beta_\mathrm{ani}},\kappa_\mathrm{ext}|\boldsymbol{d_\mathrm{ACS}},\sigma,\boldsymbol{d_\mathrm{env}},\boldsymbol{\Delta
t})$ is the posterior given all the available data and $P_2 =
P(\boldsymbol{D},\gamma',\theta_E,\gamma_\mathrm{ext},\boldsymbol{\eta},\boldsymbol{\beta_\mathrm{ani}},\kappa_\mathrm{ext}|\boldsymbol{d_\mathrm{ACS}},\boldsymbol{\Delta
t})$ is the posterior given only the lens image and time-delay data. Bayes' theorem yields
\begin{equation}
\frac{P_1}{P_2} \propto P(\sigma,\boldsymbol{d_\mathrm{env}}|\boldsymbol{D},\gamma',\theta_E,\gamma_\mathrm{ext},\boldsymbol{\eta},\boldsymbol{\beta_\mathrm{ani}},\kappa_\mathrm{ext}).
\end{equation}
This shows that when calculating the posterior distribution of
$\boldsymbol{D}$, the lensing time-delay likelihood $P_2$ can be weighted by the
likelihood of $\sigma$ and $\boldsymbol{d_\mathrm{env}}$, and then
integrated over the rest of parameters \cite{lewis/bridle:2002, suyu/etal:2010}. \cref{fig:schem} schematically
summarizes the process.

\begin{figure}[H]
\centering
\includegraphics[width=16cm]{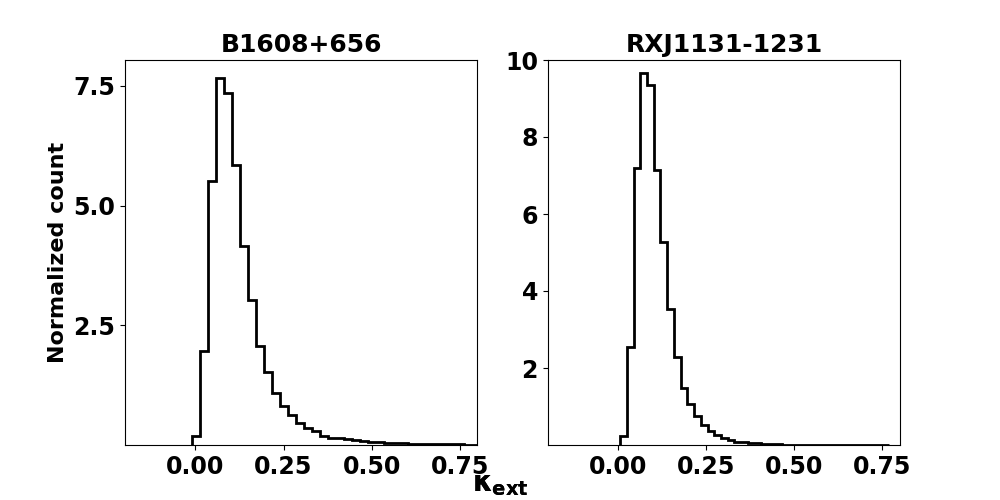}
\\(A)$~~~~~~~~~~~~~~~~~~~~~~~~~~~~~~~~~~~~~~~~~~~~~~~~~~~~~~~~~$(B)
\caption{\textbf{The distribution of external convergence for each lens.}}
\label{fig:ext_conv}
\end{figure}

As shown above, the kinematics likelihood can be calculated separately
and incorporated into the lensing likelihood. We give flat priors to
$D_\mathrm{d}$ and $D_\mathrm{ds}/D_\mathrm{s}$ and use PDF of
$\kappa_\mathrm{ext}$ from \cite{hilbert/etal:2009, suyu/etal:2012, suyu/etal:2010}
\cref{fig:ext_conv} as a prior $P(\kappa_\mathrm{ext})$.
The posterior PDFs of the angular diameter distances
are independent of $P(\kappa_\mathrm{ext})$ \cite{jee/etal:2015a}.

\section{Combining Two Anisotropy Models}
\label{sec:comb}
We calculate the posterior distribution of $D_\mathrm{d}$ using two
anisotropy models, OM and TPE using a wide range of uninformative
uniform priors on anisotropy parameters. However, as there is
no spatially-resolved dynamics data of these lenses available to constrain the anisotropy of the
system, one model is not preferred over the other. For RXJ1131$-$1231, we tested two different mass models: composite and power-law, and as the two models are degenerate, both models are equally preferred. We thus conclude that
both the anisotropy models (and the mass models) are equally likely, and infer cosmology from the
two (four for RXJ1131$-$1231) distributions combined. We created equal-weight posterior distributions for all models of interest (two anisotropy models for B1608+656 and two mass and two anisotropy models for RXJ1131$-$1231) and concatenated them. From this combined posterior distribution, we fitted a shifted log normal distribution with three parameters $\sigma_{\rm sft}$, $\lambda_{\rm sft}$ and $\mu_{\rm sft}$,
\begin{equation}
P(x, \sigma_{\rm sft}, \lambda_{\rm sft}, \mu_{\rm sft}) = \frac{1}{\sqrt{2\pi}\sigma_{\rm sft} (x-\lambda_{\rm sft})} \exp\left[-\frac{\log^2((x-\lambda_{\rm sft})/\mu_{\rm sft})}{2 \sigma_{\rm sft}^2}\right],
\end{equation}
where 
\begin{equation}
(\sigma_{\rm sft},\lambda_{\rm sft},\mu_{\rm sft}) = (0.1836, 334.2, 894.9)
\end{equation}
for B1608+656 and 
\begin{equation}
(\sigma_{\rm sft},\lambda_{\rm sft},\mu_{\rm sft})    = (0.2460, 222.5, 590.8)
\end{equation} for RXJ1131$-$1231, with $x = D_{\rm d}/{\rm Mpc}$. From this shifted log normal distribution, the median, 16$^\mathrm{th}$ and 84$^\mathrm{th}$ percentile values of the original posterior distribution were recovered with precisions $<1\%$ for B1608+656 and $<2\%$ for RXJ1131$-$1231.  
\section{Combining the JLA type Ia SNe and the Lensing Distances}
\label{sec:sne}
We adopt a distance modulus model \cite{betoule/etal:2014} in the JLA cosmological analysis, and combine it with the two lensing distances inferred. \\
As type Ia SNe are standardizable candles, cosmological analysis \cite{betoule/etal:2014} empirically assumes that SNe in same color, shape and host galaxy environment have the same luminosity over all redshifts. Under these assumptions, the distance modulus $\mu_{\rm dist}$ to each SNe can be calculated following \cref{eq:distance_modulus} once four nuisance parameters $\alpha$, $\beta$, $M_{\rm B}$ and $\Delta M$ are determined; $\alpha$ scales the stretch of the light curve in time-domain, $\beta$ scales the color at the peak of the light curve, $M_{\rm B}$ is the absolute $B$ band magnitude of the SNe at the peak of the light curve, and $\Delta M$ characterizes the change in the peak absolute magnitude as a function of the stellar mass of the host galaxy. In particular,
\begin{equation}
    \mu_{\rm dist} = 5 \log_{10}(d_L/10 \mathrm{pc}) = m_B^*-(M_{\rm B}-\alpha \times X_1 - \beta \times C),
    \label{eq:distance_modulus}
\end{equation}
where $d_L$ is the luminosity distance, $m_B^*$ is the observed peak magnitude of the SNe in the rest-frame B band, $X_1$ is the light-curve stretch parameter in time and $C$ is the color of a SN at its maximum brightness. Thus, once the nuisance parameters are constrained, one can measure the luminosity distance to a SN if the standardized light-curve is available.\\
The shape and color parameters ($X_1$ and $C$, respectively) of individual SNe are available from the SALT2 light-curve model, as well as the  covariance matrix of these light-curve parameters and software to compute the likelihood of the cosmological parameters and the nuisance parameters using all the SNe in the JLA dataset \cite{betoule/etal:2014}.\\
For the lensing distance likelihood, we assume that the two lensing distances that we measured are independent of each other as they are well separated in redshift/location, and there is no physical reason for correlation. For each lens, we combine the posterior distributions that we obtained from importance sampling the lens models with different anisotropy models (and mass models in case of RXJ1131$-$1231), and use the mean and the standard deviation of the combined distributions as representing distance to the lens and its uncertainty. We use these lensing $D_{\rm d}$ measurements to compute the log-likelihood of the cosmological parameters and add it to SNe log-likelihood to combine lensing and SNe, as the lensing distances and SNe distances are independent.\\
By combining the two experiments at the likelihood level, we simultaneously calibrate SNe at all redshifts marginalising  over the nuisance parameters and obtain constraints on the cosmological parameters.

\begin{figure}
\begin{center}
\includegraphics[height=17cm]{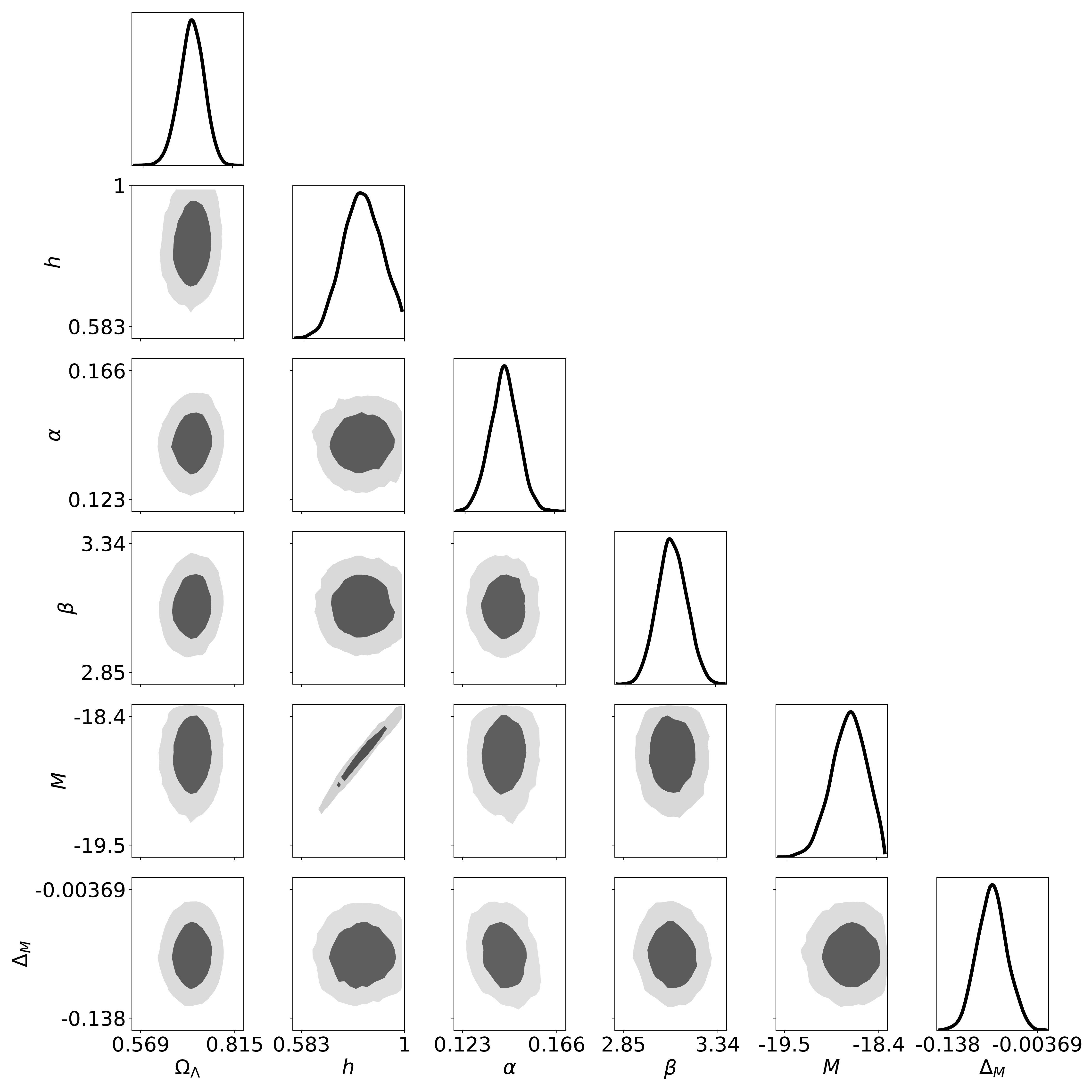}
\caption{\label{fig:param_const} \textbf{Cosmological and type Ia SNe nuisance parameters constraints for flat $\Lambda$CDM model.} One- and two-dimensional posterior distributions of
the cosmological and type Ia SNe nuisance parameters constrained using the angular diameter distances from two lenses and JLA, for flat $\Lambda$CDM model. The dark and light shaded area in two-dimensional contours are for 68\% and 95\% Confidence Level.}
\end{center}
\end{figure}

We tabulate the mean cosmological parameter constraints on six cosmological models in \cref{tb:cosmo} and the maximum likelihood parameters of the same models in \cref{tb:cosmo_bf}. The constraints on $H_0$ are quite robust
against changes in the cosmological model, despite that the constraints
on the other cosmological parameters vary substantially (e.g. strongly non-flat cosmology, which can be easily ruled out by other data sets), a feature of the inverse distance ladder method \cite{aubourg/etal:2015,cuesta/etal:2014}. 
We show the two-dimensional marginalized
joint constraints for a flat $\Lambda$CDM model in \cref{fig:param_const}.

\begin{table}
\centering
\caption{\textbf{Summary of the cosmological parameters for six cosmological
 models tested.} The parameters here are the mean of the marginalized likelihoods. $\Omega_{\Lambda}$ and $\Omega_k$ are the dark energy density
 and curvature parameters, $h$ is the dimensionless Hubble constant
 defined as $H_0=100~h~{\rm km\,s^{-1}\,Mpc^{-1}}$, and $w_0$ and $w_a$ are the dark
 energy equation of state parameters. We impose a prior that $\Omega_\Lambda>0$.}
\begin{tabular}{cccccc}
  \toprule[1pt]
  \head{Model}&\head{$\Omega_{\Lambda}$} & \head{$\Omega_k$} & \head{$h$} & \head{$w_0$} & \head{$w_a$}\\
   \midrule[1.5pt]
   f$\Lambda$CDM&0.703$^{+0.036}_{-0.033}$ & $\equiv0$ & 0.824$^{+0.084}_{-0.083}$ & $\equiv-1$ & $\equiv0$\\
   \midrule[0.5pt]
   f$w$CDM&0.776$^{+0.115}_{-0.134}$& $\equiv0$ &0.827$^{+0.085}_{-0.093}$ &$-0.89^{+0.30}_{-0.13}$& $\equiv0$\\
   \midrule[0.5pt]
   f$w_a$CDM&0.748$^{+0.113}_{-0.134}$ & $\equiv0$ &0.826$^{+0.082}_{-0.094}$&$-0.90^{+0.30}_{-0.14}$& $-14.3^{+15.5}_{-5.5}$ \\
   \midrule[0.5pt]
   nf$\Lambda$CDM&0.562$^{+0.160}_{-0.157}$ &0.23$^{+0.26}_{-0.25}$ &0.826$^{+0.085}_{-0.091}$& $\equiv-1$ & $\equiv0$ \\
   \midrule[0.5pt]
   nf$w$CDM&0.276$^{+0.003}_{-0.187}$&0.61$^{+0.29}_{-0.05}$&0.834$^{+0.084}_{-0.098}$&$-5.6^{+5.2}_{-1.3}$& $\equiv0$\\
   \midrule[0.5pt]
   nf$w_a$CDM&0.263$^{+0.015}_{-0.166}$&0.63$^{+0.26}_{-0.07}$&0.832$^{+0.084}_{-0.093}$& $-4.5^{+4.1}_{-9.9}$&$-4.1^{+11.6}_{-10.4}$\\
  \bottomrule[1pt]
\end{tabular}
\label{tb:cosmo}
\end{table}

\begin{table}
\caption{\label{tb:cosmo_bf} \textbf{Summary of the maximum likelihood parameters for six cosmological
 models tested.} Notation is the same as Table S1.}
\centering
\begin{tabular}{cccccc}
  \toprule[1pt]
  \head{Model}&\head{$\Omega_{\Lambda}$} & \head{$\Omega_k$} & \head{$h$} & \head{$w_0$} & \head{$w_a$}\\
   \midrule[1.5pt]
   f$\Lambda$CDM&0.695 & $\equiv0$ & 0.811 & $\equiv-1$ & $\equiv0$\\
   \midrule[0.5pt]
   f$w$CDM&0.822& $\equiv0$ &0.817 &$-0.8$& $\equiv0$\\
   \midrule[0.5pt]
   f$w_a$CDM&0.812 & $\equiv0$ &0.828&$-0.8$& $-1.53$ \\
   \midrule[0.5pt]
   nf$\Lambda$CDM&0.530 &0.29 &0.802& $\equiv-1$ & $\equiv0$ \\
   \midrule[0.5pt]
   nf$w$CDM&0.152&0.86&0.865&$-17.5$& $\equiv0$\\
   \midrule[0.5pt]
   nf$w_a$CDM&0.172&0.81&0.849& $-7.2$&23.1\\
  \bottomrule[1pt]
\end{tabular}
\end{table}

